\documentclass[aps,prc,letterpaper,11pt,twoside,tightenlines,nofootinbib,showpacs,preprint]{revtex4-1}
\usepackage{epsfig}
\usepackage{subfigure}
\usepackage{amsmath}
\begin{document}
\arraycolsep1.5pt
\newcommand{\Ima}{\textrm{Im}}
\newcommand{\Rea}{\textrm{Re}}
\newcommand{\mev}{\textrm{ MeV}}
\newcommand{\gev}{\textrm{ GeV}}

\def\bra#1#2#3#4#5{_{(\ell_M={#4}/2,\,\ell_B={#5})}\big\langle S_{c\bar
    c}={#1},\, {\cal L}=\frac{{#2}}{2}\,;  J=\frac{{#3}}{2}|}
\def\bracc#1#2#3#4{_{(\ell_M=0,\,\ell_B=\frac{{#4}}{2})}\big\langle S_{c\bar
    c}={#1},\, {\cal L}=\frac{{#2}}{2}\,; J=\frac{{#3}}{2}|}
\def\ket#1#2#3#4#5{| S_{c\bar c}={#1},\, {\cal L}=\frac{{#2}}{2}\,;J=\frac{{#3}}{2}\big\rangle_{(\ell_M={#4}/2,\,\ell_B={#5})}}
\def\ketcc#1#2#3#4{| S_{c\bar c}={#1},\, {\cal L}=\frac{{#2}}{2}\,; J=\frac{{#3}}{2}\big\rangle_{(\ell_M=0,\,\ell_B=\frac{{#4}}{2})}}

\title{Hidden beauty baryon states in the local hidden gauge approach with heavy quark spin symmetry}

\author{C. W. Xiao and E. Oset}

\affiliation{
Departamento de F\'{\i}sica Te\'orica and IFIC, Centro Mixto Universidad \\de Valencia-CSIC, Institutos de Investigaci\'on de Paterna, Apartado 22085, 46071 Valencia, Spain
}

\date{\today}

\begin{abstract}

Using a coupled channel unitary approach, combining the heavy quark spin symmetry and the dynamics of the local hidden gauge, we investigate the meson-baryon interaction with hidden beauty and obtain several new states of $N^*$ around 11 GeV. We consider the basis of states $\eta_b N$, $\Upsilon N$, $B \Lambda_b$, $B \Sigma_b$, $B^* \Lambda_b$, $B^* \Sigma_b$, $B^* \Sigma_b^*$ and find four basic bound states which correspond to $B \Sigma_b$, $B \Sigma_b^*$, $B^* \Sigma_b$  and $B^* \Sigma_b^*$, decaying mostly into $\eta_b N$ and $\Upsilon N$ and with a binding energy about $50-130$ MeV with respect to the thresholds of the corresponding channel. All of them have isospin $I=1/2$, and there are no bound states or resonances in $I=3/2$. The $B \Sigma_b$ state appears in $J=1/2$, the $B \Sigma_b^*$ in $J=3/2$, the $B^* \Sigma_b$ appears nearly degenerate in $J=1/2, ~3/2$ and the $B^* \Sigma_b^*$ appears nearly degenerate in $J=1/2, ~3/2, ~5/2$. These states have a width from $6 \sim 45 \mev$ except for the one in $J=5/2$ which has zero width since it cannot decay into any of the states of the basis chosen.
 
\end{abstract}

\maketitle

\section{Introduction}

The advent of the quark model allowed to put much order into the plethora of elementary particles and resonances. During much time the quark model with three quarks for the baryons and $q \bar{q}$ for the mesons was the standard picture. The panorama evolved with the time, giving rise to more complex configurations and even pentaquarks and heptaquarks for the baryons or tetraquarks for the mesons (see recent reviews in \cite{Crede:2013kia,Klempt:2009pi,Klempt:2000ud}). Yet, the growing number of states and accumulation of data has revealed that there are even more complex structures. A turning point in this direction was the introduction of chiral dynamics to study meson-meson and meson-baryon interaction, which produced as a surprise that many known resonances, called dynamically generated, were coming as a natural consequence of the hadron-hadron interaction, much is the way as the deuteron appears as a simple bound state of a proton and a neutron. By evaluating the kernel from the chiral Lagrangian \cite{Weinberg:1978kz,Gasser:1984gg,Meissner:1993ah,Pich:1995bw,Ecker:1994gg,Bernard:1995dp}, the coupled channel approach implementing unitarity explains successfully the experimental data and properties of the light scalar mesons \cite{Oller:1997ti,Oller:1998hw,Dai:2011bs}, $a_0(980),~f_0(980),~\sigma,~\kappa$, etc, and the light baryons \cite{Kaiser:1995eg,angels,Inoue:2001ip,Jido:2003cb,GarciaRecio:2002td,Hyodo:2002pk}, two $\Lambda(1405),~\Lambda(1670),~N^*(1535),~\Delta(1620)$, etc. Extrapolation of this dynamics to the charm sector has also produced many meson states, as the $D^*_{s0}(2317),~D^*_0(2400),~X(3700),~X(3872)$, etc \cite{kolo,hofmann,Guo:2006fu,Gamermann:2006nm,danielax}, as well as baryon states like the $\Lambda_c(2595)$ \cite{hofmann2,Mizutani:2006vq,Tolos:2007vh}. When the interaction involves vector mesons, the natural extrapolation is the use of the local hidden gauge Lagrangians \cite{hidden1,hidden2,hidden3,hidden4}, which include pseudoscalar and vector mesons, incorporating chiral symmetry and providing the interaction of the vector mesons, or vector mesons with baryons (see recent review \cite{review}).

Under the SU(3) symmetry, the local hidden gauge Lagrangian with the coupled channel unitary approach can explain the structure, properties and dynamics of many states which are confirmed in the Particle Data Group (PDG) \cite{pdg2012}. With this formalism, the $\rho \rho$ interaction is studied in \cite{raquelrho}, and provides a natural explanation of the meson states $f_0(1370)$ and $f_2(1270)$ and obtains the masses and widths of the two particles in fair agreement with experimental results. Along the same line, the work of \cite{gengvec} successfully finds 11 states in the vector meson-vector meson interaction, five of which are identified as $f_0(1370)$, $f_0(1710)$, $f_2(1270)$, $f_2′(1525)$, and $K_2^*(1430)$, reported in the PDG and regarded as molecular states, and the other ones are predictions. An extension of this method to the case of the nonet of vectors interacting with the decuplet baryons is done in \cite{kolobar,sarkar}, dynamically generating some resonances found in PDG. Turning to the vector nonet-baryon octet interactions, there are results obtained about $J^P = 1/2^-, ~3/2^-$ particles in the work of \cite{angelsvec}. Extension of these ideas to incorporate simultaneously pseudoscalar mesons, vector mesons and baryons is done in \cite{javier,kanchan1,kanchan2,kanchan3}. The meson-meson interaction with charm is studied in \cite{raquelhideko,xyz,raqueltanya}, which dynamically generates the particles $D_2^*(2460), ~X(3940), ~Z(3940), ~X(4160)$ and $D_{s2}^*(2573)$. For the meson-baryon interaction, the works \cite{wuprl,wuprc} extrapolate the hidden gauge formalism with the coupled channel approach to the hidden charm sector, and dynamically generate some narrow $N^*$ and $\Lambda^*$ resonances around $4.3\gev$, not listed in the PDG. Analogously, the work of \cite{wuzou} extends this later formalism to the hidden beauty sector and also predicts several $N^*$ and $\Lambda^*$ states with narrow width and energies around $11\gev$.

On the other hand, in the heavy quark sectors there is another symmetry, heavy quark spin-flavour symmetry as stated in \cite{hqss1,hqss2,hqss3,hqss4}, or heavy quark spin symmetry (HQSS) as described in \cite{hqssguo}, which predicts an $\eta'_c f_0(980)$ bound state, suggested as the spin-doublet partner of the $Y(4660)$ theoretically proposed as a $\psi' f_0(980)$ bound state in \cite{Guo:2008zg}. Incorporating the HQSS and the effective field theory, the charmed meson-antimeson system is investigated in \cite{Nieves:2011zz,Nieves:2012tt,HidalgoDuque:2012pq,HidalgoDuque:2012ej}, predicting six hidden charm states as HQSS partners of the $D \bar{D}^*$ bound state, $X(3872)$, two of which are assumed to be $X(3915)$, a $D^* \bar{D}^*$ molecular state, and $Y(4140)$,  a $D^*_s \bar{D}^*_s$ molecular state. Under the SU(8) spin-flavour symmetry requirement, within the framework of the unitary coupled channel approach, the meson-baryon interactions are studied in \cite{GarciaRecio:2008dp,Gamermann:2010zz,Romanets:2012hm,GarciaRecio:2012db,Garcia-Recio:2013gaa}, and some charmed and strange baryon resonances are produced dynamically in their theoretical models. A step forward in this direction is given in \cite{xiaojuan} combining the local hidden gauge formalism and HQSS, and using a the unitary coupled channel method, making a prediction of four hidden charm states with relatively small widths. In the present work, we extrapolate this later approach to the hidden beauty sector. We also propose a natural way to regularize the loops which removes ambiguities encountered in other works \cite{wuzou}, and find 7 dynamically generated states with hidden beauty, with small widths, some of them degenerate in  spin, which present a challenge for experimental investigation.

\section{HQSS and local hidden gauge Formalism}
\label{sechqss}

Following the work of \cite{xiaojuan}, we extrapolate the formalism to the hidden beauty sector by just changing the $\bar{D}$ meson to a $B$ meson and $c$-quark to $b$-quark. Therefore we can study baryons with hidden beauty with isospin $I=1/2,\ 3/2$, and spin $J=1/2,\ 3/2,\ 5/2$. We take as coupled channels states with $\eta_b,\ \Upsilon$ and a $N$ or a $\Delta$, and states with $B,\ B^*$ and $\Lambda_b,\ \Sigma_b$ or $\Sigma_b^*$. For the different $I,\ J$ quantum numbers we have the following space states.\\

1) $J=1/2,\ I=1/2$

$\quad \eta_b N,\ \Upsilon N,\ B \Lambda_b,\ B \Sigma_b,\ B^* \Lambda_b,\ B^* \Sigma_b,\ B^* \Sigma_b^*$.\\

2) $J=1/2,\ I=3/2$

$\quad \Upsilon \Delta,\ B \Sigma_b,\ B^* \Sigma_b,\ B^* \Sigma_b^*$.\\

3) $J=3/2,\ I=1/2$

$\quad \Upsilon N,\ B^* \Lambda_b,\ B^* \Sigma_b,\ B \Sigma_b^*,\ B^* \Sigma_b^*$.\\

4) $J=3/2,\ I=3/2$

$\quad \eta_b \Delta,\ \Upsilon \Delta,\ B^* \Sigma_b,\ B \Sigma_b^*,\ B^* \Sigma_b^*$.\\

5) $J=5/2,\ I=1/2$

$\quad B^* \Sigma_b^*$.\\

6) $J=5/2,\ I=3/2$

$\quad \Upsilon \Delta,\ B^* \Sigma_b^*$.\\

We have 17 orthogonal states in the physical basis. In order to take into account the HQSS it is interesting to use the heavy quark basis in which the spins are rearranged such as to combine the spin of the $b \bar{b}$ quarks into $S_{b \bar{b}}$ since the matrix elements do not depend on this spin. One clarifies the HQSS in terms of $\vec{S}_{b \bar{b}}, ~\vec{\cal L}$ (the spin of the three light quarks) and $\vec{J}$ (the total spin of the system). The conservation of $\vec{S}_{b \bar{b}}$ and $\vec{J}$ leads to the conservation of $\vec{\cal L} = \vec{J} - \vec{S}_{b \bar{b}}$ and then in the HQSS basis the matrix elements fulfil
\begin{equation}
\begin{split}
& _{(\ell_M',\ell_B')}\big\langle S'_{b\bar b},\, {\cal L}'; J',\, \alpha'|H^{QCD}|
S_{b\bar b},\, {\cal L}; J,\, \alpha \big \rangle_{(\ell_M,\ell_B)} \\
= & \;
\delta_{\alpha \alpha'}\delta_{JJ'}\delta_{S'_{b\bar b}S_{b\bar b} }
\delta_{{\cal L}{\cal L}'}  \big\langle \ell_M'\ell_B' {\cal L};
\alpha ||H^{QCD}  || \ell_M\ell_B {\cal L}; \alpha \big\rangle, \label{eq:hqs}
\end{split}
\end{equation}
where $\ell_B, ~\ell_M$ are the spins of the light quarks in the baryon and meson respectively under consideration (see \cite{xiaojuan} for details). By changing from the physical meson-baryon basis to the one of the HQSS states and evaluating the transition matrix elements between the physical states, using Eq. \eqref{eq:hqs} one obtains the same structure as found in \cite{xiaojuan} which we reproduce below for the explicit states of the present problem.
\begin{itemize}
\item $J=1/2$, $I=1/2$
\[
\left. \phantom{(}
\begin{array}{ccccccc}
\phantom{ \sqrt{\frac{2}{3}} \text{ $\mu_{13}$}} & \phantom{\frac{\sqrt{2} \text{ $\mu_{13}$}}{3} } & \phantom{\sqrt{\frac{2}{3}} \text{  $\mu_{23}$}} & \phantom{\frac{1}{3} \sqrt{\frac{2}{3}} (\text{$\mu_3$}-\text{$\lambda_2 $})} &
   \phantom{\frac{\sqrt{2} \text{$\mu_{23}$}}{3}} & \phantom{\frac{1}{9} \sqrt{2}
   (\text{$\mu_3$}-\text{$\lambda_2 $})} & \phantom{\frac{1}{9}
   (\text{$\lambda_2 $}+8 \text{$\mu_3$})}\\
\eta_b N & \Upsilon N &  B \Lambda_b &  B \Sigma_b &  B^* \Lambda_b
  &  B^* \Sigma_b &  B^* \Sigma^*_b 
\end{array}
\right. \phantom{)_{I=1/2}}
\]
\begin{equation}
\left(
\begin{array}{ccccccc}
 \text{$\mu_1$} & 0 & \frac{\text{$\mu_{12}$}}{2} &
 \frac{\text{$\mu_{13}$}}{2} & \frac{\sqrt{3} \text{$\mu_{12}$}}{2} &
 -\frac{\text{$\mu_{13}$}}{2 \sqrt{3}} & \sqrt{\frac{2}{3}}
 \text{$\mu_{13}$} \\ \\
 0 & \text{$\mu_1$} & \frac{\sqrt{3} \text{$\mu_{12}$}}{2} &
 -\frac{\text{$\mu_{13}$}}{2 \sqrt{3}} & -\frac{\text{$\mu_{12}$}}{2}
 & \frac{5 \text{$\mu_{13}$}}{6} & \frac{\sqrt{2}
   \text{$\mu_{13}$}}{3} \\ \\
 \frac{\text{$\mu_{12}$}}{2} & \frac{\sqrt{3} \text{$\mu_{12}$}}{2} &
 \text{$\mu_2$} & 0 & 0 & \frac{\text{$\mu_{23}$}}{\sqrt{3}} &
 \sqrt{\frac{2}{3}} \text{$\mu_{23}$} \\ \\
 \frac{\text{$\mu_{13}$}}{2} & -\frac{\text{$\mu_{13}$}}{2 \sqrt{3}} & 0 & \frac{1}{3} (2 \text{$\lambda_2 $}+\text{$\mu_3$}) & \frac{\text{$\mu_{23}$}}{\sqrt{3}} & \frac{2 (\text{$\lambda_2$}-\text{$\mu_3$})}{3 \sqrt{3}} & \frac{1}{3} \sqrt{\frac{2}{3}}
 (\text{$\mu_3$}-\text{$\lambda_2 $}) \\ \\
 \frac{\sqrt{3} \text{$\mu_{12}$}}{2} & -\frac{\text{$\mu_{12}$}}{2} & 0 & \frac{\text{$\mu_{23}$}}{\sqrt{3}} & \text{$\mu_2$} & -\frac{2 \text{$\mu_{23}$}}{3} & \frac{\sqrt{2} \text{$\mu_{23}$}}{3} \\ \\
 -\frac{\text{$\mu_{13}$}}{2 \sqrt{3}} & \frac{5 \text{$\mu_{13}$}}{6} & \frac{\text{$\mu_{23}$}}{\sqrt{3}} & \frac{2 (\text{$\lambda_2 $}-\text{$\mu_3$})}{3 \sqrt{3}} & -\frac{2 \text{$\mu_{23}$}}{3} & \frac{1}{9} (2 \text{$\lambda_2 $}+7 \text{$\mu_3$}) &
 \frac{1}{9} \sqrt{2} (\text{$\mu_3$}-\text{$\lambda_2 $}) \\ \\
 \sqrt{\frac{2}{3}} \text{$\mu_{13}$ } & \frac{\sqrt{2} \text{$\mu_{13}$}}{3}\; & \sqrt{\frac{2}{3}} \text{$\mu_{23}$ } & \frac{1}{3} \sqrt{\frac{2}{3}} (\text{$\mu_3$}-\text{$\lambda_2 $})\; &
   \frac{\sqrt{2} \text{$\mu_{23}$}}{3}\;\; & \frac{1}{9} \sqrt{2}
   (\text{$\mu_3$}-\text{$\lambda_2 $}) & \frac{1}{9}
   (\text{$\lambda_2 $}+8 \text{$\mu_3$}) \\ \\
\end{array}
\right)_{ I=1/2}
\label{eq:ji11}
\end{equation}

\item $J=1/2$, $I=3/2$
\[
\left. \phantom{(}
\begin{array}{cccc}
 \phantom{-\frac{\text{$\lambda_{12}$}}{3}} & \phantom{\frac{1}{3} \sqrt{\frac{2}{3}} (\text{$\mu_3$}-\text{$\lambda_2$})} & \phantom{\frac{1}{9} \sqrt{2} (\text{$\mu_3$}-\text{$\lambda_2$})} & \phantom{\frac{1}{9}
   (\text{$\lambda_2$}+8 \text{$\mu_3$})} \\
\Upsilon \Delta &  B \Sigma_b &  B^* \Sigma_b &  B^* \Sigma^*_b 
\end{array}
\right. \phantom{)_{I=3/2}}
\]
\begin{equation}
\left(
\begin{array}{cccc}
 \text{$\lambda_1$} & \sqrt{\frac{2}{3}} \text{$\lambda_{12}$} &
 \frac{\sqrt{2} \text{$\lambda_{12}$}}{3} &
 -\frac{\text{$\lambda_{12}$}}{3} \\ \\
 \sqrt{\frac{2}{3}} \text{$\lambda_{12}$} & \frac{1}{3} (2 \text{$\lambda_2$}+\text{$\mu_3$}) & \frac{2 (\text{$\lambda_2$}-\text{$\mu_3$})}{3 \sqrt{3}} & \frac{1}{3}
   \sqrt{\frac{2}{3}} (\text{$\mu_3$}-\text{$\lambda_2$}) \\ \\
 \frac{\sqrt{2} \text{$\lambda_{12}$}}{3} & \frac{2 (\text{$\lambda_2$}-\text{$\mu_3$})}{3 \sqrt{3}} & \frac{1}{9} (2 \text{$\lambda_2$}+7 \text{$\mu_3$}) & \frac{1}{9} \sqrt{2}
   (\text{$\mu_3$}-\text{$\lambda_2$})\\ \\
 -\frac{\text{$\lambda_{12}$}}{3} & \frac{1}{3} \sqrt{\frac{2}{3}} (\text{$\mu_3$}-\text{$\lambda_2$})\;\; & \frac{1}{9} \sqrt{2} (\text{$\mu_3$}-\text{$\lambda_2$}) & \frac{1}{9}
   (\text{$\lambda_2$}+8 \text{$\mu_3$}) \\
\end{array}
\right)_{ I=3/2}
\label{eq:ji13}
\end{equation}
\item  $J=3/2$, $I=1/2$
\[
\left. \phantom{(}
\begin{array}{ccccc}
\phantom{\frac{\sqrt{5} \text{$\mu_{13}$}}{3}} & \phantom{\frac{\sqrt{5}
   \text{$\mu_{23}$}}{3}} & \phantom{\frac{1}{9} \sqrt{5}
 (\text{$\mu_3$}-\text{$\lambda_2 $})} & \phantom{\frac{1}{3} \sqrt{\frac{5}{3}} 
(\text{$\lambda_2$}-\text{$\mu_3$})} & \phantom{\frac{1}{9} (4 \text{$\lambda_2 $}+5 \text{$\mu_3$})}\\
 \Upsilon N &  B^* \Lambda_b &  B^* \Sigma_b 
&  B \Sigma^*_b  &  B^* \Sigma^*_b 
\end{array}
\right. \phantom{)_{I=1/2}}
\]

\begin{equation}
\left(
\begin{array}{ccccc}
 \text{$\mu_1$} & \text{$\mu_{12}$} & \frac{\text{$\mu_{13}$}}{3} & -\frac{\text{$\mu_{13}$}}{\sqrt{3}} & \frac{\sqrt{5} \text{$\mu_{13}$}}{3} \\\\
 \text{$\mu_{12}$} & \text{$\mu_2$} & \frac{\text{$\mu_{23}$}}{3} & -\frac{\text{$\mu_{23}$}}{\sqrt{3}} & \frac{\sqrt{5} \text{$\mu_{23}$}}{3} \\\\
 \frac{\text{$\mu_{13}$}}{3} & \frac{\text{$\mu_{23}$}}{3} & \frac{1}{9} (8 \text{$\lambda_2 $}+\text{$\mu_3$}) & \frac{\text{$\lambda_2 $}-\text{$\mu_3$}}{3 \sqrt{3}} & \frac{1}{9}
   \sqrt{5} (\text{$\mu_3$}-\text{$\lambda_2 $}) \\\\
 -\frac{\text{$\mu_{13}$}}{\sqrt{3}} & -\frac{\text{$\mu_{23}$}}{\sqrt{3}} & \frac{\text{$\lambda_2 $}-\text{$\mu_3$}}{3 \sqrt{3}} & \frac{1}{3} (2 \text{$\lambda_2 $}+\text{$\mu_3$}) &
   \frac{1}{3} \sqrt{\frac{5}{3}} (\text{$\lambda_2 $}-\text{$\mu_3$}) \\\\
 \frac{\sqrt{5} \text{$\mu_{13}$}}{3}\; & \frac{\sqrt{5}
   \text{$\mu_{23}$}}{3}\; & \frac{1}{9} \sqrt{5}
 (\text{$\mu_3$}-\text{$\lambda_2 $})\; & \frac{1}{3} \sqrt{\frac{5}{3}} 
(\text{$\lambda_2$}-\text{$\mu_3$})\; & \frac{1}{9} (4 \text{$\lambda_2 $}+5 \text{$\mu_3$}) \\
\end{array}
\right)_{I=1/2}
\label{eq:ji31}
\end{equation}

\item  $J=3/2$, $I=3/2$
\[
\left. \phantom{(}
\begin{array}{ccccc}
\phantom{\frac{1}{2} \sqrt{\frac{5}{3}} \text{$\lambda_{12} $}} & \phantom{\frac{1}{2} \sqrt{\frac{5}{3}} \text{$\lambda_{12} $}} & \phantom{\frac{1}{9} \sqrt{5} (\text{$\mu_3$}-\text{$\lambda_2 $})} & \phantom{\frac{1}{3} \sqrt{\frac{5}{3}}
   (\text{$\lambda_2 $}-\text{$\mu_3$})} & \phantom{\frac{1}{9} (4 \text{$\lambda_2 $}+5 \text{$\mu_3$})} \\
\eta_b \Delta & \Upsilon \Delta &  B^* \Sigma_b 
&  B \Sigma^*_b  &  B^* \Sigma^*_b 
\end{array}
\right. \phantom{)_{I=3/2}}
\]
\begin{equation}
\left(
\begin{array}{ccccc}
 \text{$\lambda_1 $} & 0 & -\frac{\text{$\lambda_{12} $}}{\sqrt{3}} & \frac{\text{$\lambda_{12} $}}{2} & \frac{1}{2} \sqrt{\frac{5}{3}} \text{$\lambda_{12} $} \\\\
 0 & \text{$\lambda_1 $} & \frac{\sqrt{5} \text{$\lambda_{12} $}}{3} & \frac{1}{2} \sqrt{\frac{5}{3}} \text{$\lambda_{12} $} & \frac{\text{$\lambda_{12} $}}{6} \\\\
 -\frac{\text{$\lambda_{12} $}}{\sqrt{3}} & \frac{\sqrt{5} \text{$\lambda_{12} $}}{3} & \frac{1}{9} (8 \text{$\lambda_2 $}+\text{$\mu_3 $}) & \frac{\text{$\lambda_2 $}-\text{$\mu_3$}}{3
   \sqrt{3}} & \frac{1}{9} \sqrt{5} (\text{$\mu_3$}-\text{$\lambda_2 $}) \\\\
 \frac{\text{$\lambda_{12} $}}{2} & \frac{1}{2} \sqrt{\frac{5}{3}} \text{$\lambda_{12} $} & \frac{\text{$\lambda_2 $}-\text{$\mu_3$}}{3 \sqrt{3}} & \frac{1}{3} (2 \text{$\lambda_2$}+\text{$\mu_3$}) & \frac{1}{3} \sqrt{\frac{5}{3}} (\text{$\lambda_2 $}-\text{$\mu_3$}) \\\\
 \frac{1}{2} \sqrt{\frac{5}{3}} \text{$\lambda_{12} $} & \frac{\text{$\lambda_{12} $}}{6} & \frac{1}{9} \sqrt{5} (\text{$\mu_3$}-\text{$\lambda_2 $})\;\; & \frac{1}{3} \sqrt{\frac{5}{3}}
   (\text{$\lambda_2 $}-\text{$\mu_3$})\; & \frac{1}{9} (4 \text{$\lambda_2 $}+5 \text{$\mu_3$}) \\
\end{array}
\right)_{I=3/2}
\label{eq:ji33}
\end{equation}

\item  $J=5/2$, $I=1/2$
\[
\left. \phantom{(}
\begin{array}{c}
\phantom{\text{$\lambda_2$}}\\
 B^* \Sigma^*_b 
\end{array}
\right. \phantom{)_{I=1/2}}
\]
\begin{equation}
\left(
\begin{array}{c}
 \text{$\lambda_2$} \\
\end{array}
\right)_{I=1/2}
\label{eq:ji51}
\end{equation}

\item  $J=5/2$, $I=3/2$
\[
\left. \phantom{(}
\begin{array}{cc}
\phantom{\text{$\lambda_{12}$}} &\phantom{\text{$\lambda_{12}$}} \\
 \Upsilon \Delta \quad & B^* \Sigma^*_b 
\end{array}
\right. \phantom{)_{I=3/2}}
\]
\begin{equation}
\left(
\begin{array}{cc}
 \text{$\lambda_1$} & \text{$\lambda_{12}$} \\\\
 \text{ $\lambda_{12}$  } & \text{  $\lambda_2$ } \\
\end{array}
\right)_{I=3/2}
\label{eq:ji53}
\end{equation}

\end{itemize}
where the coefficients $\mu_{i}^I$, $\mu_{ij}^I$ ($i,j=1,2,3$) and $\lambda_{m}^I$, $\lambda_{mn}^I$ ($m,n=1,2$) are the nine unknown low energy constants of HQSS, which depend on isospin and can be related using $SU(3)$ flavour symmetry. The values of these coefficients are also depended on the used model. Following the results of \cite{xiaojuan}, which are determined by the local hidden gauge formalism, analogously we extrapolate the local hidden gauge formalism to the beauty sector as done in \cite{wuzou}.

\begin{figure}[tb]
\epsfig{file=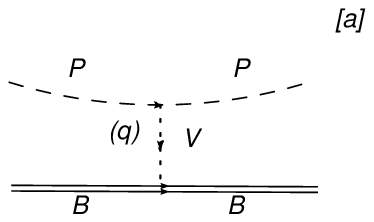, width=7cm} \epsfig{file=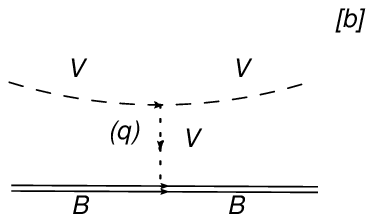, width=7cm}
\caption{Diagrams for interaction of pseudoscalar [a] or vector [b] mesons with the octet or decuplet of baryons.}\label{fig:f1}
\end{figure}

In the formalism of the local hidden gauge, the Lagrangians involving the exchanged vector mesons are given by
\begin{eqnarray}
{\cal L}_{VVV} &=& ig ~\langle [V_\nu,\partial_{\mu}V_\nu]V^{\mu}\rangle,\\
{\cal L}_{PPV} &=& -ig ~\langle [P,\partial_{\mu}P]V^{\mu}\rangle,\\
{\cal L}_{BBV} &=&
g \left( \langle \bar{B} \gamma_{\mu} [V^{\mu}, B] \rangle + \langle \bar{B} \gamma_{\mu} B \rangle \langle V^{\mu} \rangle \right),
\end{eqnarray}
where $g=m_V/(2f)$ with $f=93$~MeV the pion decay constant and taking $m_V = m_\rho$. The magnitude $V_\mu$ is the SU(4) matrix of the vectors of the meson 15-plet + singlet, $P$ the SU(4) matrix of the pseudoscalar fields, and $B$ stands for the baryon fields in SU(4) as done in \cite{wuprl,wuprc}. Starting from these Lagrangians, the $PB \to PB$ and $VB \to VB$ interaction can be shown using the Feynman diagrams by exchanging a vector meson between the pseudoscalar or the vector meson and the baryon, as depicted in Fig. \ref{fig:f1}. Note that since the mesons are of the type $u \bar{b}$, etc, then one is exchanging light vectors of type $u \bar{u}$, etc, and the heavy quarks are spectators. Note also that the possible exchange of $b \bar{b}~(\Upsilon)$ is strongly suppressed because of the heavy $\Upsilon$ mass. Thus, one can see there one of the characteristics of the HQSS symmetry where this interaction is also independent of the flavor of the heavy quarks, since there are just spectators.

Using the local hidden gauge approach, one can evaluate the interaction potentials of SU(4), shown in Fig. \ref{fig:f1}, which has this form
\begin{equation}
V_{i j}= - C_{i j} \, \frac{1}{4 f^2} \, (k^0 + k'^0),
\label{eq:vij}
\end{equation}
where $k^0, k'^0$ are the energies of the incoming and outgoing mesons (for the vector mesons, we have ignored the factor $\vec{\epsilon} ~\vec{\epsilon}~'$, following the discussion in \cite{xiaojuan}), and $C_{ij}$ are coefficients given there. Thus, combining these matrix elements with the HQSS requirements, for the matrix elements of Eq. \eqref{eq:ji11} in the $J = 1/2,~ I = 1/2$ sector, we obtain the values for the parameters of the low energy constants for $J = 1/2,~ I = 1/2$, analogously to the ones in \cite{xiaojuan}. Thus, we obtain
\begin{equation}
\begin{split}
\mu_2 &= \frac{1}{4f^2} (k^0 + k'^0),\\
\mu_3 &= -\frac{1}{4f^2} (k^0 + k'^0),\\
\mu_{12} &= -\sqrt{6}\ \frac{m_\rho^2}{p^2_{B^*} - m^2_{B^*}}\ \frac{1}{4f^2}\ (k^0 + k'^0),\\
\mu_1 &= 0,\\
\mu_{23} &= 0,\\
\lambda_2 &= \mu_3,\\
\mu_{13} &= -\mu_{12}.
\label{eq:ji11fi}
\end{split}
\end{equation}
where $p_{B^*}$ is the four momentum of $B^*$ in the $VVV$ or $PPV$ vertex (which will be discussed later). Thus, $\mu_{12}$ is small because of the much heavier $B^*$ exchanged. But we keep it since this term is the only one that allows the scattering $\eta_b N \to \eta_b N$ ($\Upsilon N \to \Upsilon N$) through intermediate inelastic states.

Similarly, the matrix of Eq. \eqref{eq:ji13} for $J=1/2$, $I=3/2$ is given by
\begin{equation}
\begin{split}
\lambda_{12} &= 3\sqrt{3}\ \frac{m_\rho^2}{p^2_{B^*} - m^2_{B^*}}\ \frac{1}{4f^2}\ (k^0 + k'^0),\\
\mu_3 &= 2 \frac{1}{4f^2} (k^0 + k'^0),\\
\lambda_2 &= \mu_3,\\
\lambda_1 &= 0.
\label{eq:ji13fi}
\end{split}
\end{equation}

Because the coefficients $\mu_{i}^I$, $\mu_{ij}^I$ and $\lambda_{m}^I$, $\lambda_{mn}^I$ are isospin dependent but $J$ independent, the results of Eq. \eqref{eq:ji11fi} are the same for Eq. \eqref{eq:ji31} in $J=3/2, I=1/2$ and Eq. \eqref{eq:ji51} in $J=5/2, I=1/2$. The other two $I=3/2$ matrices of Eq. \eqref{eq:ji33} and Eq. \eqref{eq:ji53} share the same parameters as Eq. \eqref{eq:ji13fi}.

\section{The coupled channel approach}
\label{secicca}

The scattering matrix is evaluated by solving the coupled channels Bethe-Salpeter equation in the on shell factorization approach of \cite{angels,ollerulf}
\begin{equation}
T = [1 - V \, G]^{-1}\, V,
\label{eq:Bethe}
\end{equation}
where the kernel $V$ has been discussed in the former section and the propagator $G$ is the loop function of a meson and a baryon, which is given by
\begin{equation}
G(s) = i \int\frac{d^{4}q}{(2\pi)^{4}}\frac{2M_{B}}{(P-q)^{2}-M^{2}_{B}+i\varepsilon}\,\frac{1}{q^{2}-M^{2}_{P}+i\varepsilon},
\label{eq:G}
\end{equation}
where $M_P, ~M_B$ are the masses of meson and baryon respectively, $q$ is the four-momentum of the meson, and $P$ is the total four-momentum of the meson and the baryon, thus, $s=P^2$. The integration for the $G$ function, Eq. \eqref{eq:G}, is logarithmically divergent. There are two methods to regularize it. One is the dimensional regularization and the analytic expression can be seen in \cite{ollerulf} with a scale $\mu$ and the subtraction constant $a(\mu)$ as free parameter, 
\begin{eqnarray}
G(s) &=&\frac{2M_{B}}{16\pi^2}\big\{a_{\mu}+\textmd{ln}\frac{M^{2}_{B}}{\mu^{2}}+\frac{M^{2}_{P}-M^{2}_{B}+s}{2s}\textmd{ln}\frac{M^{2}_{P}}{M^{2}_{B}}\nonumber\\
&&+\frac{q_{cm}}{\sqrt{s}}\big[\textmd{ln}(s-(M^{2}_{B}-M^{2}_{P})+2q_{cm}\sqrt{s})+\textmd{ln}(s+(M^{2}_{B}-M^{2}_{P})+2q_{cm}\sqrt{s})\nonumber\\
&&-\textmd{ln}(-s-(M^{2}_{B}-M^{2}_{P})+2q_{cm}\sqrt{s})-\textmd{ln}(-s+(M^{2}_{B}-M^{2}_{P})+2q_{cm}\sqrt{s})\big]\big\}\ ,
\end{eqnarray}
where $q_{cm}$ the three-momentum of the particle in the center mass frame. The other method to regularize is using a cut-off momentum performing the integration
\begin{equation}
G(s) = \int_0^{q_{max}} \frac{d^{3}\vec{q}}{(2\pi)^{3}}\frac{\omega_P+\omega_B}{2\omega_P\omega_B}\,\frac{2M_{B}}{P^{0\,2}-(\omega_P+\omega_B)^2+i\varepsilon},
\label{eq:Gco}
\end{equation}
where $\omega_P = \sqrt{\vec{q}\,^2+M_P^2},~\omega_B = \sqrt{\vec{q}\,^2+M_B^2}$, and $q_{max}$ is the cut-off of the three-momentum, the free parameter. Also the analytic formula of Eq. \eqref{eq:Gco} can be seen in \cite{Guo:2005wp,Oller:1998hw}.

\begin{figure}[tb]
\epsfig{file=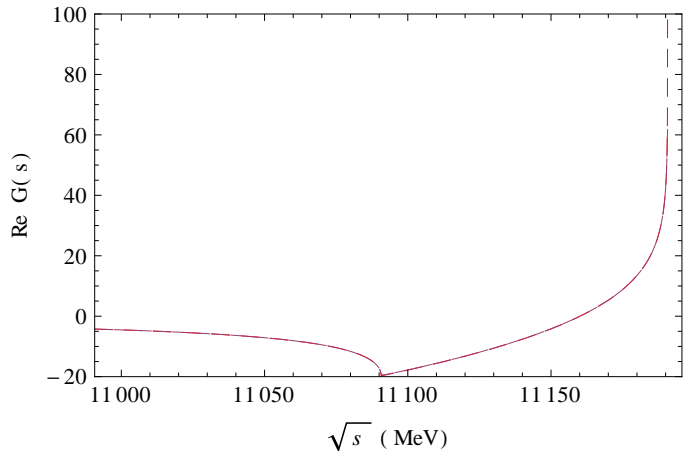, width=9cm}
\caption{Re\,$G$ as a function of $\sqrt{s}$ for $q_{max} = 745.5 \mev/c$.}\label{fig:Gqon}
\end{figure}

Normally in the low energy, the two regularization methods are compatible and there are relationships between these free parameters, $a(\mu)$, $\mu$ and $q_{max}$ (seen Eq. (52) of \cite{GarciaRecio:2010ki}). At higher energies, as discussed in \cite{wuzou}, there are large differences even not far away from threshold (see Fig. 2 of \cite{wuzou}). The cut-off method for the heavy hadrons has obvious deficiencies. Indeed, with small excitation energies where the integrand is still large, the momentum is very large. For instance, in the $B \Sigma_b$ channel, $100\mev$ of energy correspond to $745.5\mev/c$ momentum. If one takes a small cut-off of this size, as we go to excitation energies of about $100\mev$ the $G$ function blows up since there is no cancellation of the two opposite sign parts of the integrand in the principal value of the integration (see Fig. \ref{fig:Gqon}).
The physics of the problem should accommodate larger momentum than this. On the other hand, the use of $G$ in dimensional regularization has its own problems, since matching it to the cut-off formula at threshold develops positive values below threshold, leading to the unphysical generation of states with a repulsive potential when $1 - GV = 0$. The cut-off method, however does not show this pathology since $G<0$ below threshold. It is clear that one must allow larger values of $q$ inside the integral, but on doing this, the factorization of the potential due to vector meson exchange can no longer be done as one usually does in the light sector. However, what might look like a technical inconvenience works on our favor since the $\vec{q}$ dependence of the light vector meson propagator provides a physical regularization factor. Recalling that $p^0$ is small for large values of $\vec{p}$ in the heavy sector, one can take
\begin{equation}
\frac{1}{p^2-m_V^2} = \frac{1}{p^{0\,2}-\vec{p}\,^2-m_V^2} \approx \frac{1}{-\vec{p}\,^2-m_V^2} = -\frac{1}{\vec{p}\,^2+m_V^2}.
\label{eq:qexch} 
\end{equation}
For lower momentum transfers one can take the approximation, $\vec{p}\,^2 \sim 0$, and then Eq. \eqref{eq:qexch} becomes $-1/m_V^2$, which can be factorized outside the loop and give rise to the potential of Eq. \eqref{eq:vij}. In the heavy quark sector $\vec{p}$ can be larger than $m_V$ and the $\vec{p}$ dependence of Eq. \eqref{eq:qexch} must be taken into account.

We, thus, improve our formalism to solve this problem. As discussed in section VII of \cite{gamer}, we can introduce a form factor to the potential,
\begin{equation}
V(\vec{q}\,',\vec{q}\,) = \langle \vec{q}\,'|\hat{V}|\vec{q}\, \rangle \equiv v\, f(\vec{q}\,') f(\vec{q}\,).
\label{eq:Vv}
\end{equation}
Then one show in \cite{gamer} that the $T$ matrix factorizes like Eq. \eqref{eq:Vv} and one has
\begin{equation}
T(\vec{q},\vec{q}\,') = \langle \vec{q}\,|\hat{T}|\vec{q}\,' \rangle \equiv t\, f(\vec{q}\,) f(\vec{q}\,'),
\label{eq:Tt}
\end{equation}
and then the Lippmann-Schwinger equation becomes
\begin{equation}
t = [1 - v \, G]^{-1}\, v,
\label{eq:Bethe2}
\end{equation}
but now
\begin{equation}
G(s) = \int \frac{d^{3}\vec{q}}{(2\pi)^{3}} f^2(\vec{q}\,) \frac{\omega_P+\omega_B}{2\,\omega_P\,\omega_B}\,\frac{2M_{B}}{P^{0\,2}-(\omega_P+\omega_B)^2+i\varepsilon}.
\label{eq:Gco2}
\end{equation}
Once again we can put the integral equation as an algebraic equation \cite{angels}. Note that Eq. \eqref{eq:Bethe2} has the same format as Eq. \eqref{eq:Bethe}, but, the matrices $t,~v$ are defined by Eqs. \eqref{eq:Vv} and \eqref{eq:Tt}, and the loop function $G(s)$ is changed by Eq. \eqref{eq:Gco2} which absorbs a momentum dependent form factor from the factorized potential. Then, the kernel $v$ is still the same as discussed in the last section \ref{sechqss}.

\section{Application to the heavy quark sector}

In present work, we focus on the beauty sector involving much higher energy than the light quark sector, even than the charm sector. As mentioned in the former section \ref{secicca}, because of the large value of the momentum $\vec{q}$ running in the loop, we should consider the $\vec{q}$ dependence of the vector exchange. For this we use the formalism discussed in the former section.

\begin{figure}[tb]
\epsfig{file=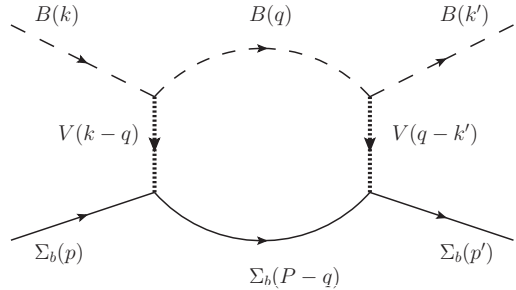, width=7cm}
\caption{Feymann diagram for the transition $B \Sigma_b \to B \Sigma_b$ with one loop.}\label{fig:bsigb}
\end{figure}

\begin{figure}[tb]
\epsfig{file=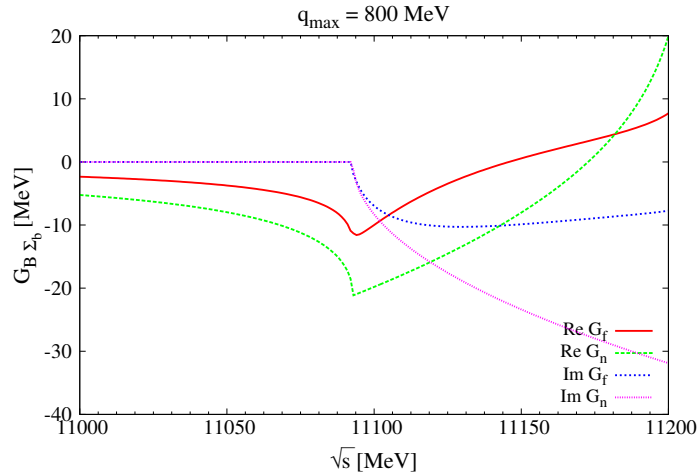, width=9cm}
\caption{The real and imaginary parts of $G$ function with Eq. \eqref{eq:Gco} ($G_n$) and Eq. \eqref{eq:Gco2} ($G_f$).}\label{fig:gloop}
\end{figure}

First, for the channels involving the light vector mesons exchange, such as $B \Sigma_b$ channel, the problem is that the potential has a factor which does not depend on $\vec{q}$ but just on $\vec{k} - \vec{q}$, as shown in Fig. \ref{fig:bsigb}.
However, we should keep in mind that while $\vec{q}$ in the loop can be larger than $m_V$, we only study states close to threshold where the external momenta are small. Thus, we have
\begin{equation}
f(\vec{k}\,)f(\vec{q}\,) \equiv \frac{m_V^2}{(\vec{k}-\vec{q}\,)^2 + m_V^2} \simeq \frac{m_V^2}{\vec{q}\,^2+m_V^2},
\end{equation}
which defines
\begin{equation}
f(\vec{q}\,) = \frac{m_V^2}{\vec{q}\,^2+m_V^2},~f(\vec{k}\,) \simeq 1.
\end{equation}
For the main potential related to the light vector meson exchange, this form factor should be incorporated into the new $G$ function, Eq. \eqref{eq:Gco2}, thus, there is a factor $f^2(\vec{q}\,)$ in the integral. With the implementation of the form factor in Eq. \eqref{eq:Gco2} the function $G$ becomes convergent. In Fig. \ref{fig:gloop}, we compare the new results for Re\,$ G$ and Im\,$ G$ with the new prescription with the sharp cut-off results with $q_{max} = 800\mev/C$ used in \cite{wuzou}. As we can see, both Re\,$ G$ and Im\,$ G$ are reduced in the new approach which leads to smaller binding of the states.

\begin{figure}[tb]
\subfigure[ ]{\label{fig:etabna}\includegraphics[scale=0.6]{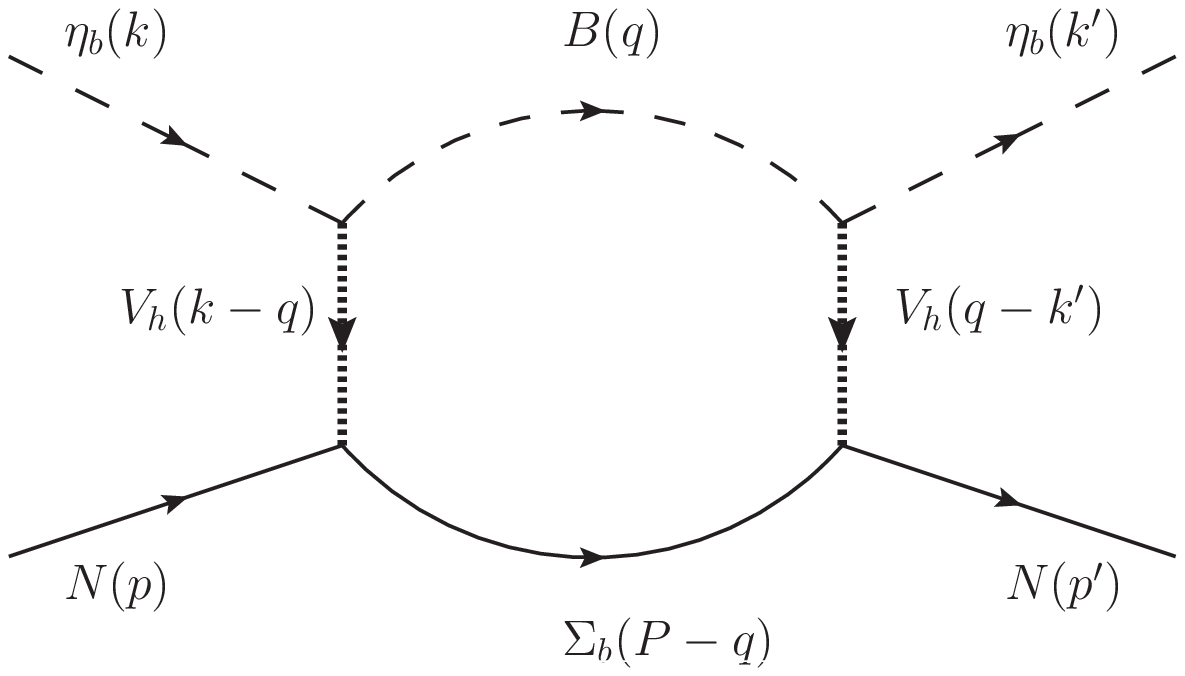}}
\subfigure[ ]{\label{fig:etabnb}\includegraphics[scale=0.6]{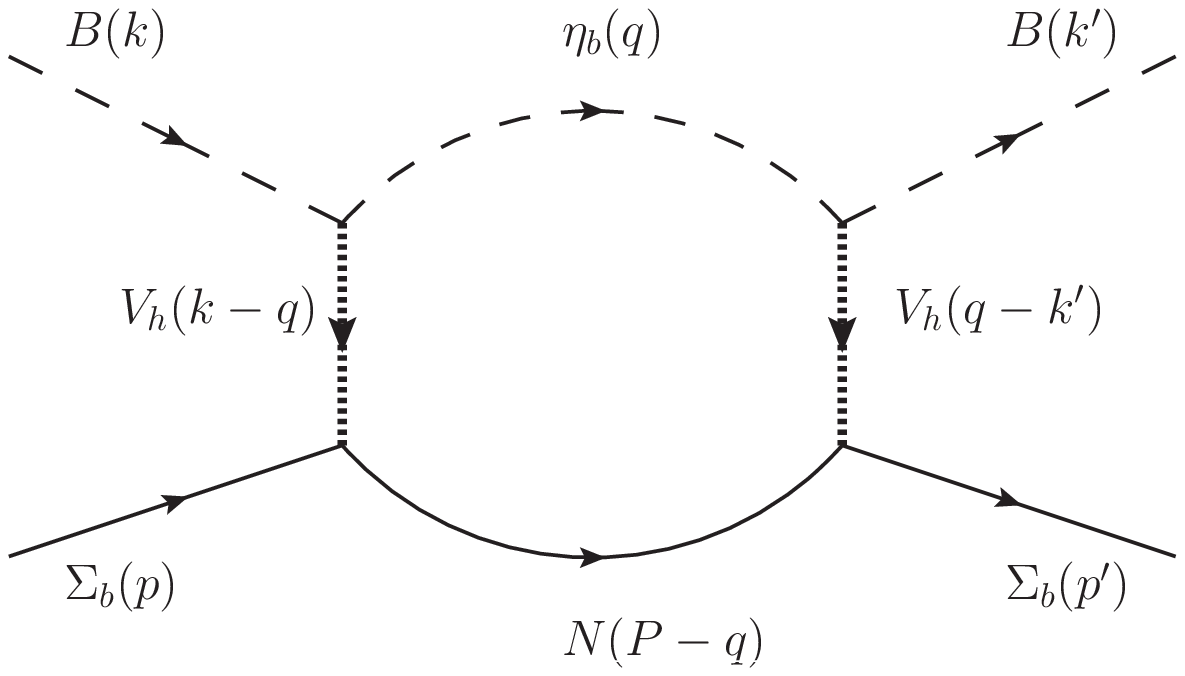}}
\caption{Diagram for the transitions coupled with $\eta_b N$ channel in the loop.}\label{fig:etabn}
\end{figure}

Second, we still face another problem when we have the transitions coupling the channels $\eta_b(\Upsilon)-N(\Delta)$, which involve the much heavier vector exchange, $B^*$. This is analogous to be $\eta_c(J/\psi)-N(\Delta)$ transition, which requires $D^*$ exchange in the charm sector. There are two cases in our coupled channel formalism, seen in Fig. \ref{fig:etabn}. 
One can find diagrams like those on Fig. \ref{fig:etabna} involving $\eta_b N \to \eta_b N$ amplitudes through the necessary intermediate states $B Y_b$ ($Y_b$ for $\Lambda_b$ or $\Sigma_b$) since the direct transition $\eta_b N \to \eta_b N$ is null (see Eq. \eqref{eq:ji13fi} $\mu_1 = 0$). Obviously this amplitude will be very small since it involves the exchange of $B^*$. We are not interested in this amplitude. The diagram of Fig. \ref{fig:etabnb} involves a term of the $B Y_b \to B Y_b$ amplitude through $\eta_b N$ intermediate state. This amplitude is also small, but it is the part that provides the width to the $B Y_b$ states since $\eta_b N$ has smaller mass than $B Y_b$ . Hence, we keep this diagram and since the $B Y_b \to \eta_b N$ transition is small we do not need to consider more than one $\eta_b N$ loop when we evaluate the $B Y_b$ unitarized amplitude. Hence, it is easy to implement this channel by also providing a form factor $\tilde{f} (\vec{q}\,)$ when we have an intermediate $\eta_b N$ state. This form factor is also easy to implement as we discuss below.

To determine the new form factor, $\tilde{f}(\vec{q}\,)$, we should come back to the transition potential $\mu_{12}$ of Eq. \eqref{eq:ji11fi}, which takes into account the heavy $B^*$ exchange propagator,
\begin{equation}
\frac{m_V^2}{p^2_{B^*}-m^2_{B^*}}.\label{eq:pbstar}
\end{equation}
If we calculate the four momentum $p^2_{B^*}$ by taking on shell approximation, we have
\begin{equation}
p^2_{B^*} = (p_{\eta_b}-p_B)^2 \simeq m^2_{\eta_b} + m^2_B - 2 E_{\eta_b}E_B,
\label{eq:pbstaron}
\end{equation}
where the on shell energies of the particles are given by
\begin{equation}
E_{\eta_b} = \frac{s + m^2_{\eta_b} - m^2_N}{2\sqrt{s}};~~E_B = \frac{s + m^2_B - m^2_{\Sigma_b}}{2\sqrt{s}}.
\label{eq:eon}
\end{equation}
Once again, we take into account that in the loop one can exchange large momenta with small energy transfer. Therefore, we can consider that the energy is the same but there will be an off shell momentum running. Thus, we take
\begin{equation}
p^2_{B^*} = (p_{\eta_b}-p_B)^2 = (E_{\eta_b} - E_B)^2 - (\vec{p}_{\eta_b}-\vec{p}_B)^2 \simeq (E_{\eta_b} - E_B)^2 - \vec{q}\,^2,
\end{equation}
where we have taken the external momentum $\vec{p}_B \approx 0$ as before and $\vec{p}_{\eta_b} = \vec{q}$. Hence, for the transition potential of Eqs. \eqref{eq:ji11fi} and \eqref{eq:ji13fi} we shall use the on shell expression, Eqs. \eqref{eq:pbstar} and \eqref{eq:pbstaron}, as in the charm sector, but now in the $\eta_b N$ channel we should use a form factor in the loop function,
\begin{equation}
\tilde{f}(\vec{q}\,) = \frac{m^2_{B^*} - (E_{\eta_b} - E_B)^2}{m^2_{B^*} - (E_{\eta_b} - E_B)^2 + \vec{q}\,^2},
\end{equation}
where the on shell energies, $E_{\eta_b}$ and $E_B$, are given by Eq. \eqref{eq:eon}. In practice, for $E_B$ we take average masses of $b$-baryons and then have a unique form factor $\tilde{f}(\vec{q}\,)$.

\section{Results and discussion}

In our formalism we use the Bethe-Salpeter equation of Eq. \eqref{eq:Bethe2} in coupled channels to evaluate the scattering amplitudes, where the $G$ function for the meson-baryon interaction is given by Eq. \eqref{eq:Gco2}. We firstly search the resonance peak in the scattering amplitudes and then look for poles in the second Riemann sheet when there are open channels, or in the first Riemann sheet when one has stable bound states (see \cite{wuprc,luis} for details).

Let $\sqrt{s_p}$ be the complex energy where a pole appears. Close to a pole the amplitude behaves as
\begin{equation}
T_{ij}=\frac{g_{i}g_{j}}{\sqrt{s}-\sqrt{s_p}}\ .\label{eq:tgigj}
\end{equation}
where $g_i$ is the coupling of the resonance to the $i$ channel. As one can see in Eq. \eqref{eq:tgigj}, $g_{i}g_{j}$ is the residue of $T_{ij}$ at the pole. For a diagonal transitions we have
\begin{equation}
g_{i}^{2}=\lim_{\sqrt{s}\rightarrow
\sqrt{s_p}}~T_{ii}\,(\sqrt{s}-\sqrt{s_p}).\label{eq:coup}
\end{equation}
The determination of the couplings gives us an idea of the structure of the states found, since according to \cite{gamer,junko,aceti}, the couplings are related to the wave function at the origin for each channel.

Similarly to the charm sector, in all $I=3/2$ channels we have repulsive potentials as can be seen in Eq. \eqref{eq:ji13fi}. So, we should not expect any bound states or resonances.

\begin{figure}[tb]
\epsfig{file=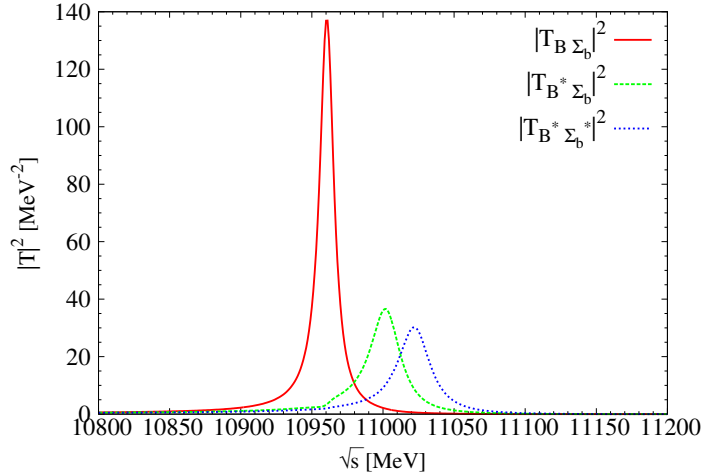, width=9cm}
\caption{The squared amplitudes of the $J=1/2,~I=1/2$ sector.}\label{fig:res11}
\end{figure}

Next we show the results for the $J=1/2,~I=1/2$ sector in Fig. \ref{fig:res11}.
There are three clear peaks with non zero width between the range $10950 \sim 11050 \mev$ in the 
squared amplitudes of $|T|^2$. These peaks are below the thresholds of $B \Sigma_b, ~B^* \Sigma_b, ~B^* \Sigma_b^*$ respectively. From Eqs. \eqref{eq:ji11} and \eqref{eq:ji11fi}, we know that the potentials of these channels are attractive, and the energy ranges where these peaks appear are reasonable. In Fig. \ref{fig:gloop}, one can see that the real parts of the loop function $G$, Eq. \eqref{eq:Gco2}, are negative below the threshold \footnote{On top of the form factors we impose an upper limit for $\vec{q}$ in the integration of $2000\mev/c$. Changes to $1500\mev/c$ or $3000\mev/c$ only lead to moderate changes on the binding of about $20\mev$, which we accept as systematic uncertainties of our approach.}. Thus these peaks are acceptable as physical ones. We look for the poles corresponding to these peaks in the second Riemann sheet, and find the poles at $(10963.04+i8.59)\mev, ~(11002.81+i19.97)\mev, ~(11023.55+i22.75)\mev$. We can see that the width of the first pole is about $17\mev$, and the last two ones have a width of about $40 \sim 45\mev$, which is three times bigger than the first one. The couplings to the various coupled channels for these poles are given in Table \ref{tab:cou11}.
\begin{table}[ht]
     \renewcommand{\arraystretch}{1.2}
\centering
\caption{The couplings of all channels corresponded certain poles in the $J=1/2,~I=1/2$ sector.} \label{tab:cou11}
\begin{tabular}{cccc cccc}
\hline\hline
\multicolumn{2}{c}{$10963.04+i8.59$}  \\
\hline
   & $\eta_b N$ & $\Upsilon N$ & $B \Lambda_b$ & $B \Sigma_b$ & $B^* \Lambda_b$ & $B^* \Sigma_b$ & $B^* \Sigma_b^*$  \\
\hline
$g_i$ & $0.78-i0.35$ & $0.44-i0.32$ & $0.00-i0.00$ & $8.52-i0.49$ & $0.03-i0.05$ & $0.39+i2.51$ & $0.04+i1.23$  \\
$|g_i|$ & $0.85$ & $0.54$ & $0.00$ & $8.54$ & $0.06$ & $2.54$ & $1.23$  \\
\hline
\multicolumn{2}{c}{$11002.81+i19.97$} \\
\hline
   & $\eta_b N$ & $\Upsilon N$ & $B \Lambda_b$ & $B \Sigma_b$ & $B^* \Lambda_b$ & $B^* \Sigma_b$ & $B^* \Sigma_b^*$  \\
\hline
$g_i$ & $0.62+i0.38$ & $1.39-0.25$ & $0.03-i0.01$ & $0.35+i1.61$ & $0.05+i0.00$ & $9.00-i1.11$ & $0.82+i1.93$  \\
$|g_i|$ & $0.73$ & $1.41$ & $0.03$ & $1.65$ & $0.05$ & $9.07$ & $2.10$  \\
\hline
\multicolumn{2}{c}{$11023.55+i22.75$} \\
\hline
   & $\eta_b N$ & $\Upsilon N$ & $B \Lambda_b$ & $B \Sigma_b$ & $B^* \Lambda_b$ & $B^* \Sigma_b$ & $B^* \Sigma_b^*$  \\
\hline
$g_i$ & $1.30-i0.21$ & $0.86+i0.13$ & $0.04+i0.00$ & $0.40+i0.43$ & $0.02-i0.01$ & $0.58+i1.60$ & $8.75-i1.19$  \\
$|g_i|$ & $1.31$ & $0.87$ & $0.04$ & $0.59$ & $0.03$ & $1.70$ & $8.83$  \\
\hline
\end{tabular}
\end{table}
From the couplings in Table \ref{tab:cou11}, the first pole, $(10963.04+i8.59)\mev$, couples mostly to $B \Sigma_b$, with a threshold of $11092.81\mev$. So, it could be considered like a $B \Sigma_b$ bound state with a binding energy about $130\mev$, which is small compared to the mass of $B \Sigma_b$. The second pole, $11002.81+i19.97$, couples most strongly to $B^* \Sigma_b$ and thus, is still bound about $136\mev$ below the $B^* \Sigma_b$ threshold, $11138.60\mev$. Finally, the third pole, $11023.55+i22.75$, couples mostly to $B^* \Sigma_b^*$. It has a binding energy of $135\mev$ with respected to the $B^* \Sigma_b^*$ threshold, $11158.80\mev$. We can see that the binding energies of the three poles are similar, close to $130\mev$.  We can also see that all the three bound states decay mostly into the open channels $\eta_b N$ and $\Upsilon N$, and couple most strongly to some other $B Y_b$ channels. Note that the former two states correspond to those reported in \cite{wuzou}, which are $(11052+i0.69)\mev$ for the $B \Sigma_b$ bound state and $(11100+i0.66)\mev$ for the $B^* \Sigma_b$ bound state. The difference in the binding energies with the results of \cite{wuzou} are at most of $90\mev$, but the uncertainties in \cite{wuzou} had a range within this magnitude. Here, the natural way to regularize the loops gives us a stronger confidence in the accuracy of the results obtained, but, as mentioned before, uncertainties of about $20\mev$ must also be accepted in our model. The width obtained in \cite{wuzou} are smaller but there are more open channels in our approach and we also do not have restrictions from using a small cut-off as used in \cite{wuzou} in some cases. Note that in the $B \Sigma_b$ decay to $\eta_b N$ the on shell momentum is about $1300\mev/c$ and will be missed if a smaller cut-off is chosen to regularized $G$. The small width obtained in \cite{wuzou} comes mostly from decay to light channels \cite{wuprl,wuprc} that we neglect here. Their results show that because of higher energy in the beauty sector, these light channels have a small influence on the two bound states decay width. In our present work, we include two open channels constrained by the HQSS, $\eta_b N$ and $\Upsilon N$, which play an important role for the the decay width. This is why we get a wider decay width.

\begin{figure}[tb]
\epsfig{file=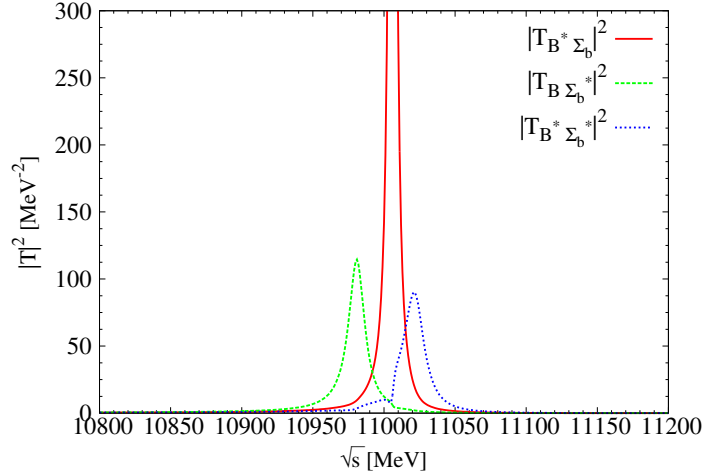, width=9cm}
\caption{The results of $|T|^2$ for the $J=3/2,~I=1/2$ sector.}\label{fig:res31}
\end{figure}

In Fig. \ref{fig:res31} we show our results for the $J=3/2,~I=1/2$ sector.
From the results of $|T|^2$, we can also see three clear peaks around the range $10950 \sim 11050 \mev$, which are about $130\mev$ below the thresholds of $B \Sigma_b^*, ~B^* \Sigma_b, ~B^* \Sigma_b^*$ respectively. The strength of the second peak is about 10 times bigger than the other two and the widths are small enough to allow the peaks to show up clearly. In the second Riemann sheet, we find the poles at $(10984.43+i9.19)\mev, ~(11007.28+i3.00)\mev, ~(11019.00+i14.80)\mev$, showing that the widths are about $18\mev, ~6\mev, ~30\mev$ respectively. We list the couplings to each coupled channel corresponding to these poles in Table \ref{tab:cou31}.
\begin{table}[ht]
     \renewcommand{\arraystretch}{1.2}
\centering
\caption{The coupling to various channels for certain poles in the $J=3/2,~I=1/2$ sector.} \label{tab:cou31}
\begin{tabular}{ccc ccc}
\hline\hline
$10984.43+i9.19$ & $\Upsilon N$ & $B^* \Lambda_b$ & $B^* \Sigma_b$ & $B \Sigma_b^*$ & $B^* \Sigma_b^*$  \\
\hline
$g_i$ & $0.93-i0.67$ & $0.04-i0.04$ & $0.06+i2.36$ & $8.79-i0.61$ & $0.66+i3.16$  \\
$|g_i|$ & $1.14$ & $0.06$ & $2.36$ & $8.81$ & $3.23$  \\
\hline
$11007.28+i3.00$ & $\Upsilon N$ & $B^* \Lambda_b$ & $B^* \Sigma_b$ & $B \Sigma_b^*$ & $B^* \Sigma_b^*$  \\
\hline
$g_i$ & $0.52-i0.41$ & $0.01+i0.09$ & $8.86-i0.39$ & $0.89+i1.39$ & $1.06+i3.54$  \\
$|g_i|$ & $0.66$ & $0.09$ & $8.87$ & $1.65$ & $3.69$  \\
\hline
$11019.00+i14.80$ & $\Upsilon N$ & $B^* \Lambda_b$ & $B^* \Sigma_b$ & $B \Sigma_b^*$ & $B^* \Sigma_b^*$  \\
\hline
$g_i$ & $1.51+i0.29$ & $0.05+i0.01$ & $1.67+i2.80$ & $0.16+i2.36$ & $9.70-i1.21$  \\
$|g_i|$ & $1.54$ & $0.05$ & $3.26$ & $2.36$ & $9.77$  \\
\hline
\end{tabular}
\end{table}
One can see from Table \ref{tab:cou31}, that the first pole, $(10984.43+i9.19)\mev$, couples most strongly to the channel $B \Sigma_b^*$ and corresponds to a $B \Sigma_b^*$ state, bound by $129\mev$ with respect to its threshold of $11113.02\mev$. The state, $(11007.28+i3.00)\mev$, corresponding to the big peak in the middle of Fig. \ref{fig:res31}, with small width, couples mostly to $B^* \Sigma_b$. Thus, it is bound by $131\mev$ with respect to the threshold of the $B^* \Sigma_b$ channel, $11138.60\mev$. The third one, $(11019.00+i14.80)\mev$, couples mostly to $B^* \Sigma_b^*$, and is bound by $140\mev$ with respect to the threshold of this channel, $11158.80\mev$. Also we can find that all the three states decay essentially into $\Upsilon N$ channel, couple very weakly to the $B^* \Lambda_b$ channel, and couple more strongly to the other channels.

\begin{figure}[tb]
\epsfig{file=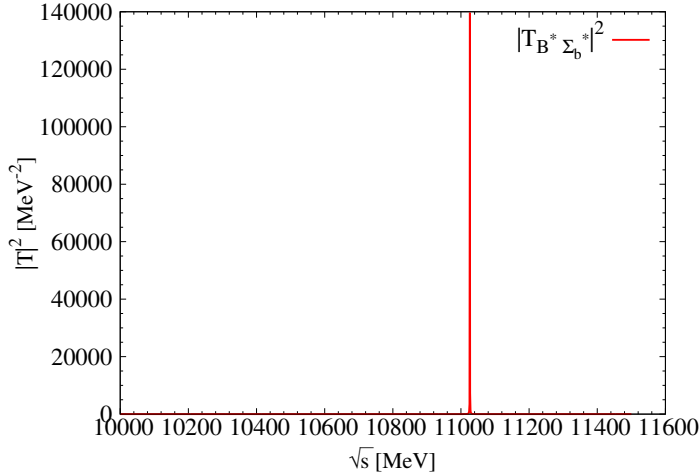, width=9cm}
\caption{The results of $|T|^2$ for the $J=5/2,~I=1/2$ sector.}\label{fig:res51}
\end{figure}

Finally, as shown in Fig. \ref{fig:res51}, we also search a new state in the $J=5/2,~I=1/2$ sector, which is a bound state of $B^* \Sigma_b^*$ around $(11026.10+i0)\mev$.
From Fig. \ref{fig:res51}, we can see that, this state has no width, as it corresponds to a single channel, $B^* \Sigma_b^*$, seen in Eq. \eqref{eq:ji51}. Then it is a bound state of this channel and has no other channels to decay. Therefore we can look for the pole in the first Riemann sheet with zero width. One can see that the state is bound by about $133\mev$ with respect to the $B^* \Sigma_b^*$ threshold.

We have seen that our procedure to regularize the loops allows sufficiently large momenta to get the imaginary part of the loops of the decay channels. Thus, we get a good estimate of the width of the states. Yet, we would like to make some estimate for the uncertainties in the masses of the states obtained. For this purpose we introduce a sharp cut off of $q_{max}=800\mev$, as suggested in \cite{wuzou}, in addition to the natural form factors from vector exchange that we have. Because of the caveat about the imaginary parts, we only look at the real parts. We observe that systematically the states are less bound. They are now bound by about $50\mev$. The experimental finding of some of the states predicted would allow us to be more refined on the regularization procedure, but for the time being we can accept the differences in the binding energies as uncertainties of our theoretical approach. We thus conclude that the states found would be bound by about $50-130\mev$ and the widths are of the order of $6-45\mev$.

\section{Conclusions}

In present work we investigate the hidden beauty sector by combining the dynamics of the local hidden gauge Lagrangians extrapolated to SU(4) with the constraints of Heavy Quark Spin Symmetry. We also benefit from the high energies of the problem and find a natural way to regularize the loops using the range provided by the light vector masses, whose exchange in the t-channel provide the source of the interaction in the local hidden gauge approach.

After our investigation, we find seven new states of $N^*$ with hidden beauty. All these states are different since they correspond to different energies or different total spin $J$.  Yet, we found that some states are bound states of the same given meson-baryon channel and appear at about the same energy but different $J$, which are analogous to those found in our former work on hidden charm. Thus, they are degenerate states that we get in $J=1/2,~3/2$ for $B^* \Sigma_b$ and $J=1/2,~3/2,~5/2$ for $B^* \Sigma_b^*$. From this perspective, we report our results as claiming that we get four bound states with about $50-130\mev$ binding and isospin $I=1/2$, corresponding to $B \Sigma_b$ with $J=1/2$, $B \Sigma_b^*$ with $J=3/2$, $B^* \Sigma_b$ degenerated with $J=1/2,~3/2$ and $B^* \Sigma_b^*$ degenerated with $J=1/2,~3/2,~5/2$. Note that the two states of $B \Sigma_b,~B^* \Sigma_b$ with $J=1/2,~I=1/2$ are consistent with the ones reported in \cite{wuzou}. Besides, we found no states in $I=3/2$. We hope that the future experiments in the BES, BELLE, FAIR and other facilities will search for these states predicted here.

\section*{Acknowledgments}  

We would like to thank J. Nieves for much help and useful discussions.
This work is partly supported by the Spanish Ministerio de Economia y Competitividad and European FEDER funds under the contract number FIS2011-28853-C02-01, and the Generalitat Valenciana in the program Prometeo, 2009/090. We acknowledge the support of the European Community-Research Infrastructure Integrating Activity Study of Strongly Interacting Matter (acronym Hadron Physics 3, Grant Agreement n. 283286) under the Seventh Framework Programme of EU.

\end{document}